\documentclass[10pt,letterpaper]{article}

\DeclareMathAlphabet{\mathpzc}{OT1}{pzc}{m}{it}

\usepackage{mathrsfs}

\usepackage{graphicx}

\usepackage{amssymb}

\usepackage[figuresright]{rotating}

\usepackage{float}

\usepackage{amsmath}

\usepackage{amsfonts}

\usepackage{theorem}
\theorembodyfont{\upshape}

\newcommand{\tableline}{\hline }

\newcommand{\be}{\begin{equation}}
\newcommand{\ee}{\end{equation}}
\newcommand{\bdm}{\begin{displaymath}}
\newcommand{\edm}{\end{displaymath}}
\newcommand{\bea}{\begin{eqnarray}}
\newcommand{\eea}{\end{eqnarray}}
\newcommand{\bse}{\small \begin{equation}}
\newcommand{\ese}{\end{equation} \normalsize}
\newcommand{\bsea}{ \small \begin{eqnarray}}
\newcommand{\esea}{\end{eqnarray} \normalsize}

\newcommand{\fig}{Fig.~\ref}
\newcommand{\tab}{Table~\ref}

\begin{document}
\title{{New results for virial
coefficients of hard spheres in \\ $D$ dimensions}}

\author{Nathan~Clisby\footnote{e-mail:\nobreak{N.Clisby@ms.unimelb.edu.au}} \\ARC Centre of Excellence for Mathematics and Statistics of Complex Systems\\ 139 Barry Street\\ The University of Melbourne, Victoria 3010\\ Australia \and Barry~M.~McCoy\footnote{e-mail:\nobreak{mccoy@insti.physics.sunysb.edu}}\\C.~N.~Yang Institute for Theoretical Physics\\ Stony Brook University\\ Stony Brook, NY 11794-3840}


\maketitle

\begin{abstract}
We  present new results for the virial coefficients $B_k$ with $k\leq 10$
for hard spheres in dimensions $D=2,\cdots ,8.$
\end{abstract}

\begin{flushright}
{\tt YITP-SB-04-57}
\end{flushright}



\section{Introduction}
\label{intro}

The low density virial
expansion of the pressure
\be \frac{P}{k_BT}=
\sum_{k=1}^{\infty}B_k \rho^{k} \hspace{0.5cm} \mathrm{with} \; B_1\equiv 1
\label{eq:virial}
\ee
for the hard sphere gas of particles of diameter $\sigma$
in $D$ dimensions defined
by the two body
potential
\be
U({\bf r}) = \left\{ \begin{array}{c} +\infty \\ 0 \end{array}
\right. \begin{array}{c} |{\bf r}|<\sigma \\ |{\bf r}|> \sigma \\
\end{array}
\label{eq:hardspherepotential}
\ee
is one of the oldest systems studied in statistical mechanics.
The problem was first studied analytically by
by van~der~Waals~\cite{vanderwaals1899a},
Boltzmann~\cite{boltzmann1899b}, and van~Laar~\cite{vanlaar1899b}
who computed the coefficients
up through $B_4.$ The computation of $B_4$ for $D=2$ was first done in
1964 by  Rowlinson~\cite{rowlinson1964a} and Hemmer~\cite{hemmer1964a}
and very recently these analytic computations for $B_4$ have been extended
to $D=4,6,8,10,12$ by the present authors
\cite{clisby2004a}, and by Lyberg~\cite{lyberg2004a} for $D=5,7,9,11$.

All other computations for the hard sphere gas are by means of computer.
This work was initiated in the
1950s for hard discs by Metropolis et al.~\cite{metropolis1953a} and
for hard spheres by Rosenbluth and
Rosenbluth~\cite{rosenbluth1954a}. Subsequently $B_6$ and $B_7$
were computed by Ree and
Hoover\cite{ree1964a,ree1964c,ree1967a} during the 1960s
and $B_8$ was computed by Janse van
Rensberg~\cite{jansevanrensburg1993a} in 1993.
Computations for $D>3$ were
initiated in 1964 by Ree and Hoover~\cite{ree1964b} who computed $B_4$
for $D=4,\cdots,11$. The coefficients $B_5$ and
$B_6$ for $D=4$ and $5$ were computed by Bishop, Masters,
and Clarke~\cite{bishop1999a} in 1999, and Bishop, Masters, and
Vlasov~\cite{bishop2004a} have recently calculated $B_7$ in
$D=4,5$ and $B_8$ in $D=4.$

In a series of papers~\cite{clisby2004b,clisby2004c,clisby2004d}
we have extended these numerical computations by computing virial
coefficients up through $B_{10}$ in $D=2,3,\cdots, 8.$
We use the method of Ree-Hoover diagrams as evaluated by
Monte Carlo integration. The details are given in ~\cite{clisby2004d}
Our results are given in
\tab{tab:numericalvirial}
in the form of $B_k/B_2^{k-1}$ where
\begin{equation}
B_2=\frac{\sigma^D \pi ^{D/2}}{2 \Gamma(1+{\frac{D}{2}})}
\label{b2}
\end{equation}

It is well known that for hard
spheres in $D$ dimensions that for some sufficiently large $k$ which
depends on $D$ that some Ree-Hoover diagrams for $B_k$ vanish
identically for geometric reasons. We give (a lower bound on)
the number of non-vanishing Ree-Hoover diagrams in
\tab{tab:diagram_number}

\begin{sidewaystable}
\centering
\scriptsize
\begin{tabular}{lllllllll}
\tableline
\tableline
\\[-1.5ex]
\multicolumn{1}{c}{$D$} &\multicolumn{1}{c}{$B_3/B_2^2$}
&\multicolumn{1}{c}{$B_4/B_2^3$} &\multicolumn{1}{c}{$B_5/B_2^4$}
&\multicolumn{1}{c}{$B_6/B_2^5$} &\multicolumn{1}{c}{$B_7/B_2^6$}
&\multicolumn{1}{c}{$B_8/B_2^7$} &\multicolumn{1}{c}{$B_9/B_2^8$}
&\multicolumn{1}{c}{$B_{10}/B_2^9$} \\[0.7ex]
\tableline
\\[-1.5ex]
2&$0.782004\cdots$&$\;\;\>0.53223180\cdots$&$   0.33355604(1)^* $&$   \;\;\>0.1988425(42) $&$0.1148728(43)$&$\;\;\>0.0649930(34)$&{$0.0362193(35)$}&$\;\;\>0.0199537(80)$\\
3&$0.625$&$\;\;\>0.2869495\cdots$&$   0.110252(1)^* $&$\;\;\>0.03888198(91)$&$0.01302354(91)$&$\;\;\>0.0041832(11)$&{$0.0013094(13)$}&$\;\;\>0.0004035(15)$\\
4&$0.506340\cdots$&$\;\;\>0.15184606\cdots$&$0.0357041(17)$&$   \;\;\>0.0077359(16)$&{$0.0014303(19)$}&$   \;\;\>0.0002888(18)$&$0.0000441(22)$&$   \;\;\>0.0000113(31)$\\
5&$0.414063\cdots$&$\;\;\>0.0759724807\cdots$&$0.0129551(13)$&$\;\;\>0.0009815(14)$&{$0.0004162(19)$}&$-0.0001120(20)$&$   0.0000747(26) $&$-0.0000492(48)$\\
6&$0.340941\cdots$&$\;\;\>0.03336314\cdots$&{$0.0075231(11) $}&$  -0.0017385(13) $&$   0.0013066(18) $&$-0.0008950(30)$&$   0.0006673(45) $&$  -0.000525(16) $\\
7&$0.282227\cdots$&$\;\;\>0.00986494662\cdots$&{$0.0070724(10) $}&$  -0.0035121(11) $&$   0.0025386(16) $&$  -0.0019937(28) $&$   0.0016869(41) $&$  -0.001514(14) $\\
8&$0.234614\cdots$&$-0.00255768\cdots$&{$0.00743092(93) $}&$  -0.0045164(11) $&$   0.0034149(15) $&$  -0.0028624(26) $&$   0.0025969(38) $&$  -0.002511(13) $\\[0.8ex]
\tableline
\end{tabular}
\normalsize
\vspace{0.5ex}
\caption[Numerical values of virial coefficients]{\centering Numerical
 values of virial coefficients. Values for $B_7$ $D>5$, $B_8$ $D>4$, $B_9$, and $B_{10}$
are new, and other values improve on published
 literature results for $B_5$ and higher except for the results for $B_5$ for $D=2,3$ which are due to Kratky~\cite{kratky1982a}.}
\label{tab:numericalvirial}
\end{sidewaystable}

\begin{table}[H]
\centering
\caption[Number of Mayer and Ree-Hoover diagrams]{\centering Number of Mayer and Ree-Hoover integrals}
\label{tab:diagram_number}
\vspace{2ex}
\small
\begin{tabular}{lccccccccc}
\tableline
\tableline
\\[-1.5ex]
& \multicolumn{9}{c}{Order}\\
& \multicolumn{1}{c}{2} & \multicolumn{1}{c}{3} & \multicolumn{1}{c}{4} & \multicolumn{1}{c}{5} & \multicolumn{1}{c}{6} & \multicolumn{1}{c}{7} &\multicolumn{1}{c}{8} & \multicolumn{1}{c}{9} & \multicolumn{1}{c}{10} \\[0.7ex]
\tableline
\\[-1.5ex]
Mayer&1 &1 &3 &10 &56 &468 &7123 & 194066 & 9743542\\
RH&1 &1 &2 &5 &23 &171 &2606 & 81564 & 4980756\\
RH/Mayer& 1& 1& 0.667& 0.500& 0.410& 0.365& 0.366 & 0.420& 0.511\\
RH, $D=1$ & 1& 1& 1& 1& 1& 1& 1& 1& 1\\
RH, $D=2$ & 1 &1 &2 &4 &15 &73& $\gtrsim$647 & $\gtrsim$8417& $\gtrsim$110529\\
RH, $D=3$ & 1 &1 &2 &5 &22 &161&$>$2334 & $>$60902& \\
RH, $D=4$ & 1 &1 &2 &5 &23 &169&$>$2556 & $>$76318 & \\
\tableline
\end{tabular}
\normalsize
\end{table}

\section{Behavior of $B_k$ for large $k$}

It is apparent from \tab{tab:numericalvirial}
that negative virial coefficients occur. This was first observed for
$B_4$ in \cite{ree1964b}.
We observe that because $B_4$ changes sign between $D=7$ and
$D=8$, $B_6$ changes sign between $D=5$ and $6$, and $B_8$ and $B_{10}$
change sign between $D=4$ and $D=5$.
This  suggests that
for large $k$ the coefficient $B_k$ may become negative 
for dimensions smaller than 5.
In particular if for $D=2$ or $D=3$ there were a value of $k$ such that
$B_k$ changed sign then approximate
equations of state obtained from the first ten virial coefficients
would be wholly inadequate to obtain the radius of convergence
of the virial series.

The most important property of the virial coefficients $B_k$ is  not
their actual numerical values for $k$ less than some finite number but
rather their asymptotic behavior as $k\rightarrow \infty$ because it
is the asymptotic value which determines the radius of convergence. Of
course no finite number of virial coefficients can give information on
the $k\rightarrow \infty$ behavior unless there is some {\it a priori}
reason to expect that the values of $k$ are already in the asymptotic
$k\rightarrow \infty$ regime.

We see in \tab{tab:diagram_number}
the dramatic effect that
the number of non-zero Ree-Hoover
integrals in two dimensions is far less than that of the number of
biconnected graphs with non-zero star content.
At $k=10$ in $D=2$ we estimate that only $0.022$ of the 
Ree-Hoover diagrams with non-zero coefficients have non-zero integrals.

The dramatic (at least in $D=2$) reduction as $k\rightarrow \infty$
in the number of non-vanishing Ree-Hoover diagrams gives a criteria for
the size of $k$ needed for $B_k$ to be in the asymptotic region.

\vspace{.1in}

{\bf Criteria 1}

The number of nonzero Ree Hoover diagrams has approached its large $k$
behavior.

\vspace{.1in}

For $k=10$ this criteria may only be fulfilled for $D=2$ and is surely not
fulfilled at all for $D\geq 5.$

Our second criteria has been presented in our previous paper
\cite{clisby2004b}

\vspace{.1in}
{\bf Criteria 2}

The loose packed diagrams (defined to be those with the number of
$\tilde f$ bonds near
their maximum value) numerically dominate $B_k$ as $k\rightarrow \infty.$

\vspace{.1in}

The validity of this criteria has been studied in detail in
\cite{clisby2004b}. Here it was seen that for $D=3$ and $k\geq 12$ the
criteria is well satisfied and that as $D$ increases the criteria is
satisfied for smaller values of $k$. However, for $D=2$ the criteria
was not satisfied even for $k$ as large as 17.

We thus conclude that there is no dimension in which both of these
criteria are simultaneously fulfilled though in $D=3$ and $D=4$ it is possible
that they both could hold for some moderate values of $k$ such as $12-14.$

\section{Ratio Analysis}

Even though we have argued that $k=10$ may not be sufficiently large
to see the true asymptotic behavior of $B_k$
it is still of
interest to determine what results are obtained if well known methods
are used to estimate the radius of convergence from the first ten virial
coefficients.

One such way of estimating the radius of convergence
is the analysis of the ratios of coefficients \cite{gaunt1980a,guttmann1989a}
where we plot
$B_k \rho_{cp}/B_{k-1}$ versus $1/k$ (and we have normalized
the virial coefficients to the density $\rho_{cp}$ of the closest
packed lattice).
The ratio extrapolated to $1/k\rightarrow 0$
will give the
radius of convergence of the series $\rho_R$ which may also expressed
in terms of the packing fraction $\eta=B_2\rho/2^{D-1}.$
If the slope of the interpolated
points approaches zero for large
$k$ then the leading singularity is a pole on the positive real axis, if the
slope is non-zero then the divergence is algebraic.

\begin{figure}[!hbt]
\centering
\includegraphics[scale=0.45,origin=c,angle=-90]{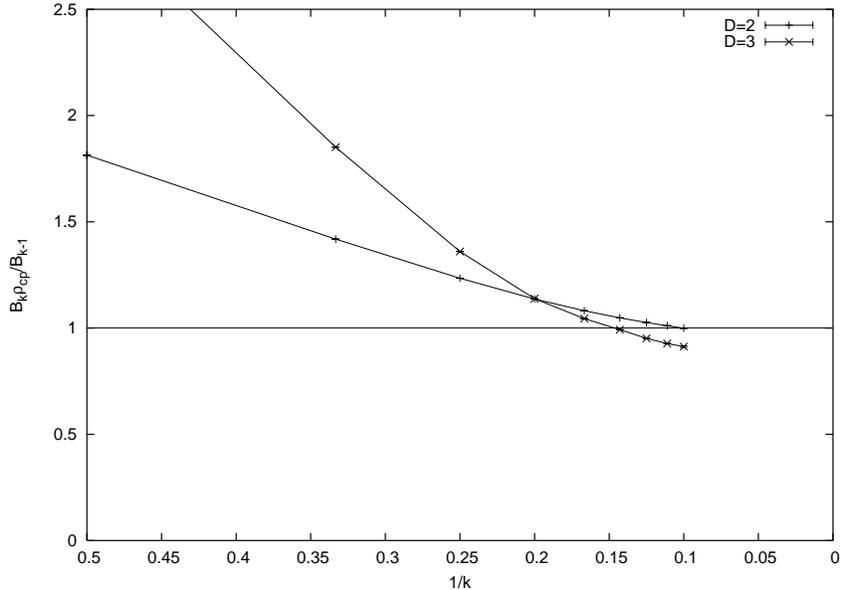}
\vspace{-2cm}
\caption[Ratio plot for virial coefficients in dimensions
 $D=2,3$]{\centering Ratio plot for
virial coefficients in dimensions $D=2,3$}
\label{graph:ratio1}
\end{figure}

The plot of $B_k\rho_{cp}/B_{k-1}$ versus $\frac{1}{k}$ for $D=2$
is given in \fig{graph:ratio1}. Here we observe smoothly falling ratios
which extrapolate to a radius of convergence greater that the closest
packed density $\rho_{cp}.$

\begin{figure}[!htb]
\centering
\includegraphics[scale=0.45,origin=c,angle=-90]{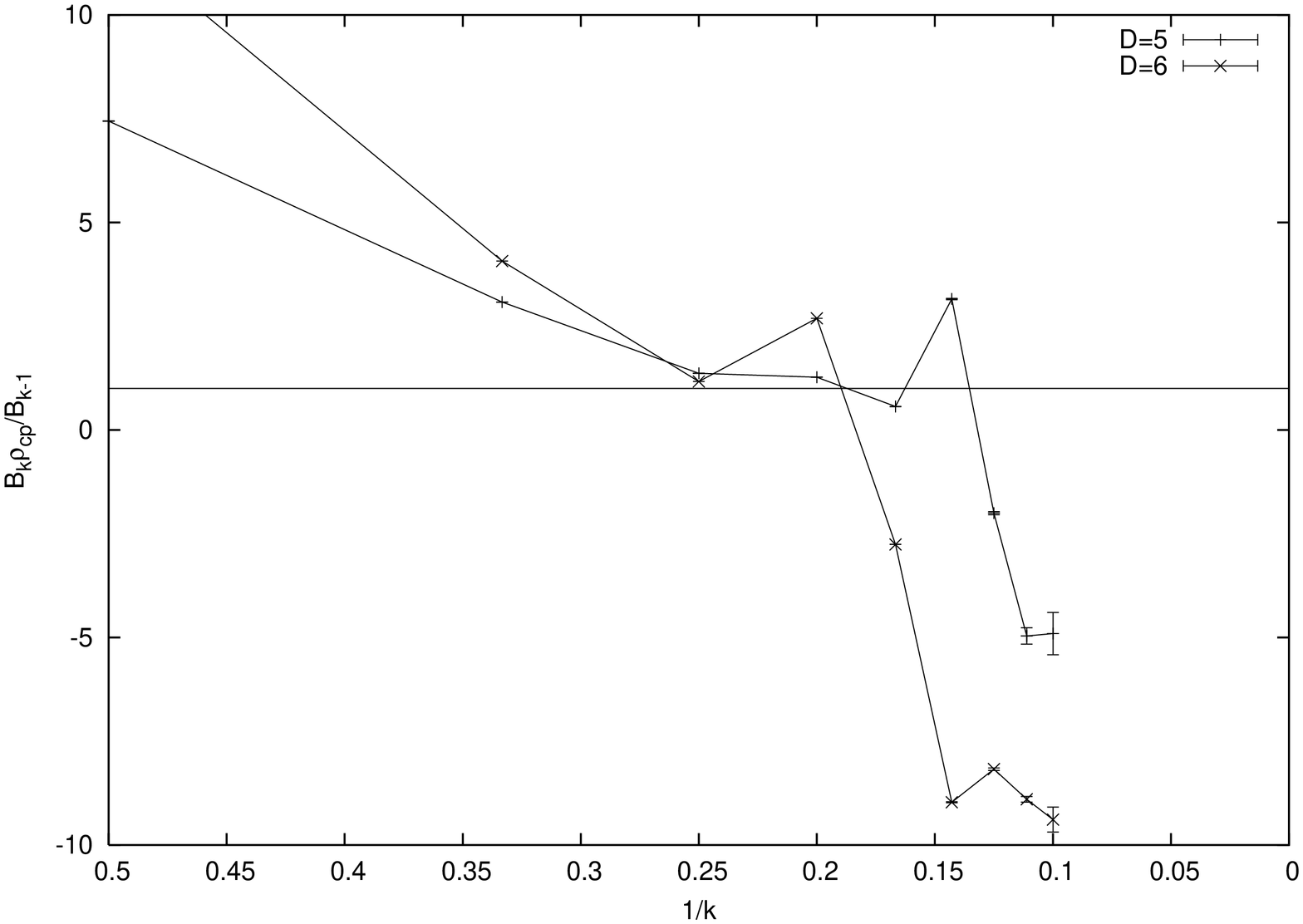}
\vspace{-2cm}
\caption[Ratio plot for virial coefficients in dimensions
 $D=5,6$]{\centering Ratio plot for
virial coefficients in dimensions $D=5,6$}
\label{graph:ratio2}
\end{figure}

\begin{figure}[!htb]
\centering
\includegraphics[scale=0.45,origin=c,angle=-90]{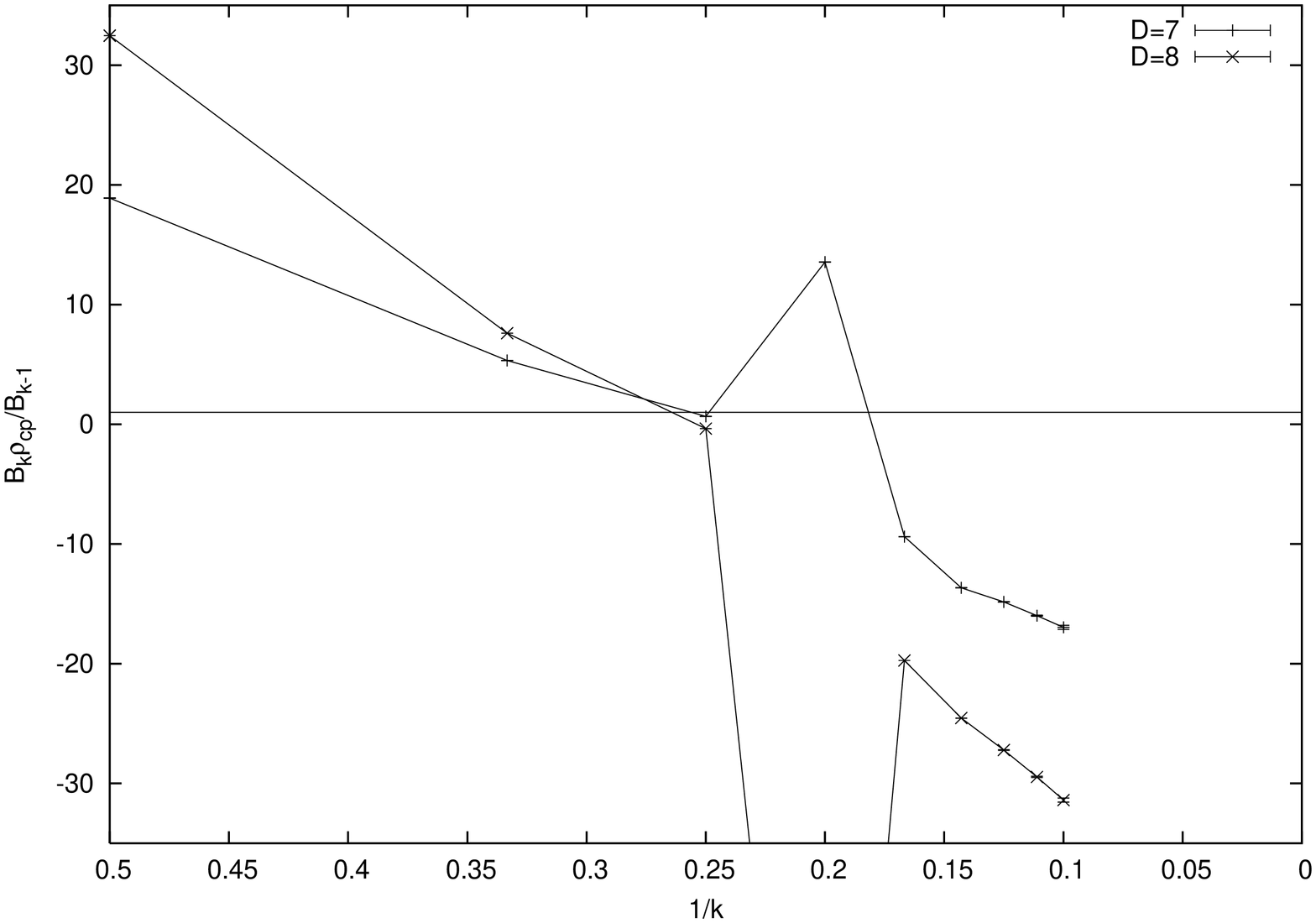}
\vspace{-2cm}
\caption[Ratio plot for virial coefficients in dimensions
 $D=7,8$]{\centering Ratio plot for
virial coefficients in dimensions $D=7,8$}
\label{graph:ratio3}
\end{figure}

We plot the ratios for $D\geq 5$
in Figures \ref{graph:ratio2} and \ref{graph:ratio3}.
In this case the first few virial coefficients are
positive, and then alternate in sign to the order
calculated. We propose
the scenario that there is a singularity on the positive real axis
that dominates the series initially, but at higher order
another singularity (or singularities) in the
complex plane or negative real axis competes with the original
singularity and hence the
new singularity must be at a smaller radius. If the leading
singularity is on the negative
real axis then the ratio plot will smoothly converge to some
negative value, otherwise
the ratios will oscillate.

\begin{figure}[!htb]
\centering
\includegraphics[scale=0.45,origin=c,angle=-90]{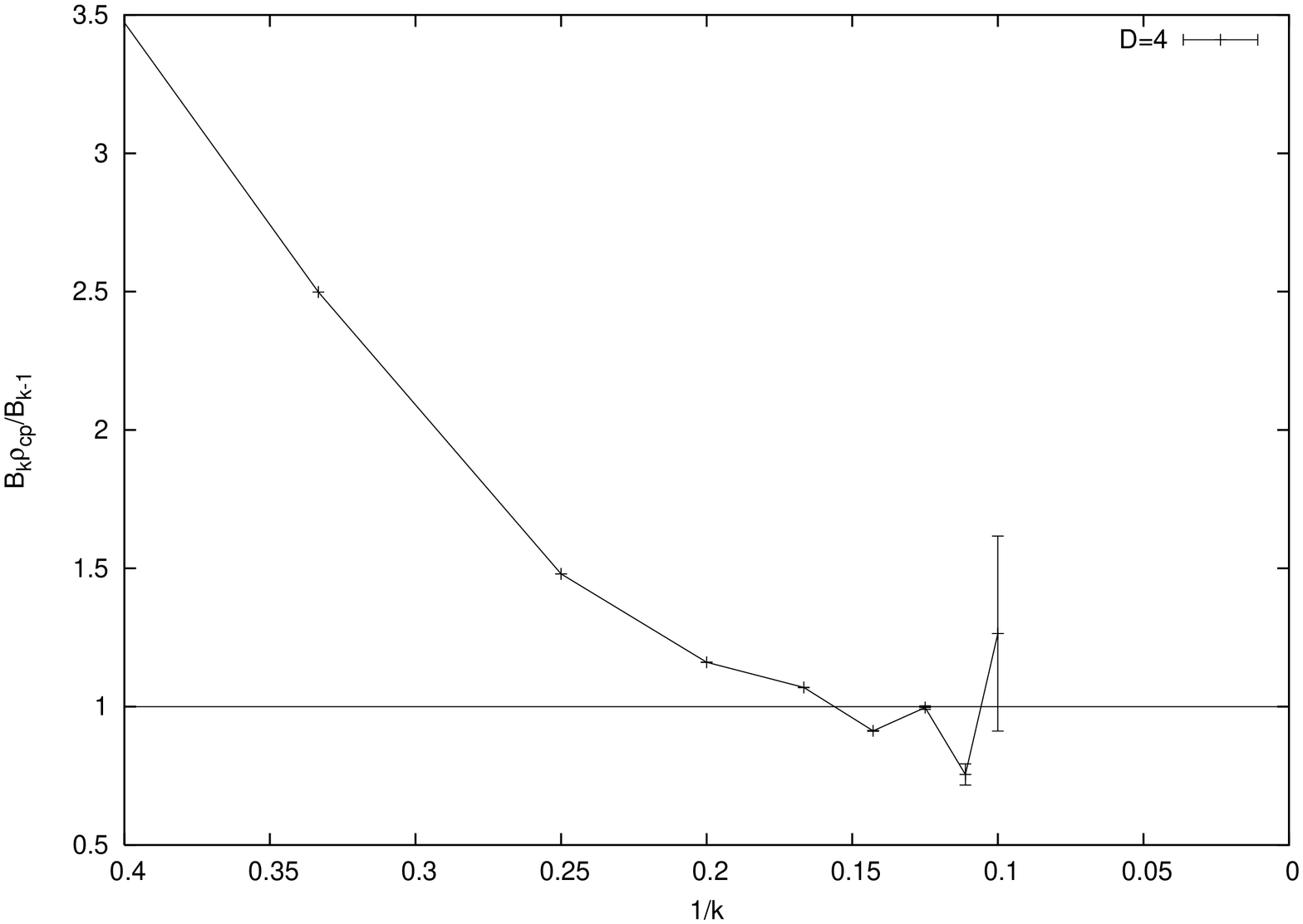}
\vspace{-2cm}
\caption[Ratio plot for virial coefficients in dimension
 $D=4$]{\centering Ratio plot for
virial coefficients in dimension $D=4$}
\label{graph:ratio4}
\end{figure}

For $D=4$ in \fig{graph:ratio4} it seems that despite
the absence of negative
virial coefficients and the poor accuracy of $B_{10}$, an oscillation
is developing in the ratio plot in exactly the same way as for
$D \ge 5$. Extrapolation of the series~\cite{clisby2004d} 
via the methods of Dlog Pad\'es
and differential approximants as explained by
Guttmann~\cite{guttmann1989a}, suggests
that negative coefficients for $D=4$ may occur for $k$
not much greater than 12.

\begin{figure}[!htb]
\centering
\includegraphics[scale=0.45,origin=c,angle=-90]{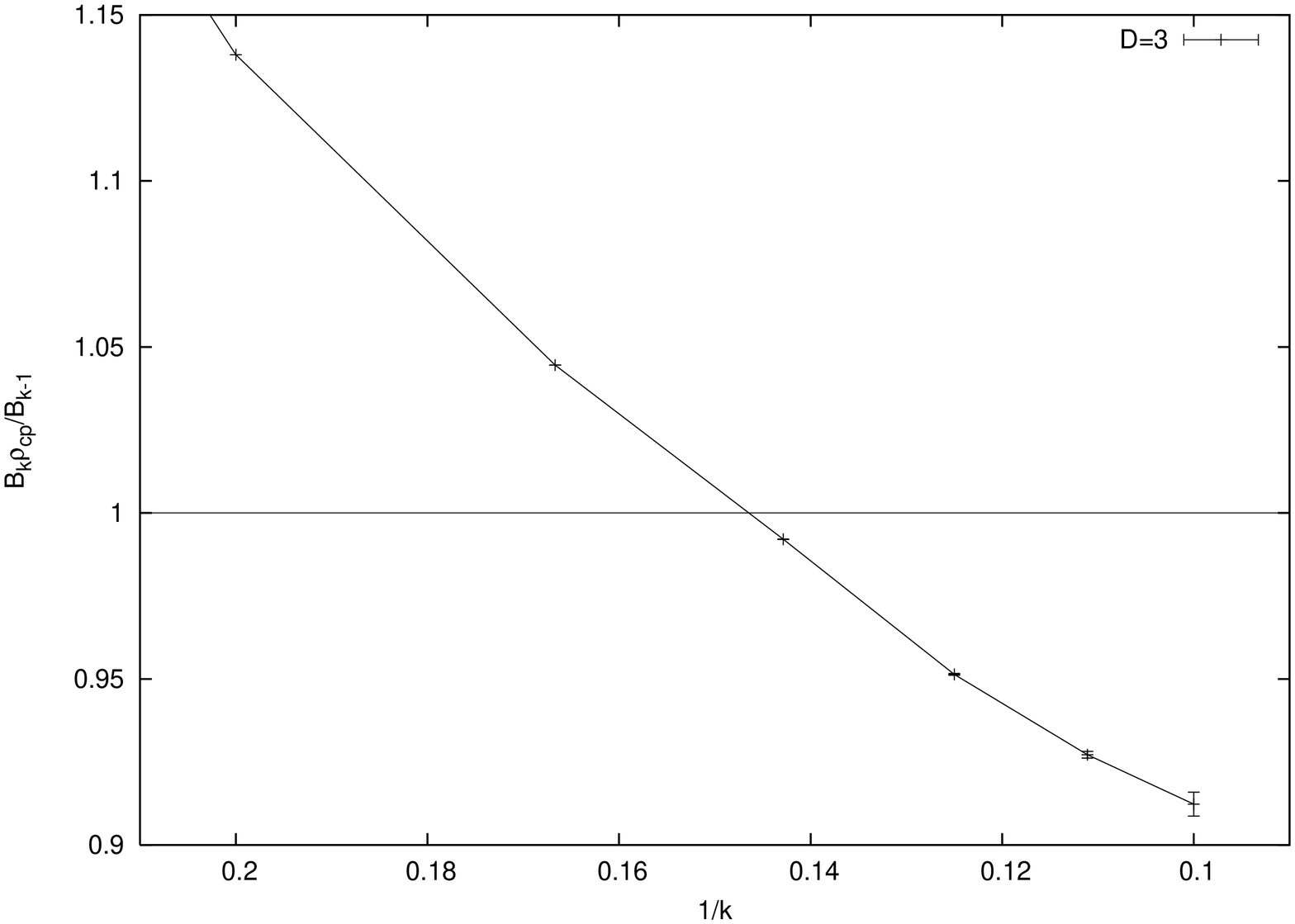}
\vspace{-2cm}
\caption[Ratio plot for virial coefficients in dimensions
 $D=3$]{\centering Ratio plot for
virial coefficients in dimension $D=3$ over small domain to show
 non-monotonic behavior of the second derivative.}
\label{graph:ratio5}
\end{figure}

The case $D=3$ is plotted in Figures \ref{graph:ratio1} and \ref{graph:ratio5}.
These ratios do not show the large oscillations of $D=4$ but close
inspection reveals that the slopes are not increasing monotonically
as they were for $D=2.$
This may indicate that the
plot for $D=3$ is displaying very small amplitude
oscillations, which will eventually result in oscillations in the sign of the
coefficients.

\section{Differential Approximants}
\label{diff}

We have analyzed the virial coefficients of \tab{tab:numericalvirial}
by use of differential approximates using the fortran program
NEWGRQD given in Guttmann~\cite{guttmann1989a}.
Our results for the leading singularity on the real positive azis in
dimensions $D=2,3,4$ are tabulated in  \tab{tab:diff1}. More detailed
analysis will appear in~\cite{clisby2004d}. The notation $L,M;N$ refers to an inhomogeneous first order
differential approximant, which is the solution of
\be z P_M(z)f^\prime(z) + Q_L(z)f(z) = R_N(z)
\ee
where the subscript denotes the order of the polynomial, and $f(z)$ is
the function that is to be approximated.

One can see from Tables \ref{tab:diff1}--\ref{tab:diff2} that
there appears to be a
singularity on the positive real axis close to the space filling
density $\eta = 1$ for dimensions $D=2,3,4$.
The kind of singularity is not so clear, for $D=2$ there seems
to be an algebraic
singularity with exponent $\phi \simeq -1.75$,
but for $D=3,4$ it is not possible
to give a good estimate for the exponent.

\begin{table}[!htb]
\centering
\caption[Singularities from differential approximants,
  $D=2,3,4$]{\centering Singularities from differential approximants
  on the positive real axis for $D=2,3,4$. Blank entries are due to
  defective approximants.}
\label{tab:diff1}
\vspace{2ex}
\begin{tabular}{ccccccc}
\tableline
\tableline
\\[-1.5ex]
\multicolumn{1}{c}{Differential}& \multicolumn{2}{c}{$D=2$}&
\multicolumn{2}{c}{$D=3$}& \multicolumn{2}{c}{$D=4$} \\
\multicolumn{1}{c}{Approximant} & \multicolumn{1}{c}{$B_2
  \rho_{sing}$} &
\multicolumn{1}{c}{$\phi$}& \multicolumn{1}{c}{$B_2\rho_{sing}$} &
\multicolumn{1}{c}{$\phi$}& \multicolumn{1}{c}{$B_2\rho_{sing}$} &
\multicolumn{1}{c}{$\phi$}\\[0.7ex]
\tableline
\\[-1.5ex]
3,3;0& 1.987&-1.790& 4.068&-2.818 & 7.995& -3.520\\
3,4;0& 1.984&-1.774& 3.830& -2.329& 6.843& -2.478\\
4,3;0& 1.984&-1.774& 3.714& -2.043& 5.551& -1.207\\
4,4;0&1.987&-1.788&3.732 & -2.090& 7.249& -2.871\\
4,5;0&1.988&-1.795&3.675 & -1.899& 6.985& -2.583\\
5,4;0&1.988&-1.795&3.720 & -2.056& 6.721& -2.229\\
2,2;1&1.946&-1.575&3.659 & -2.014& 6.888& -2.593\\
2,3;1&1.970&-1.695&3.787 & -2.246& 5.412& -2.922\\
3,2;1&1.966&-1.677&4.038 & -2.953& 8.332& -4.368\\
3,3;1&2.021&-2.076&3.811 & -2.298& 7.296& -2.920\\
3,4;1&1.981&-1.756&3.786 & -2.241& 6.764& -2.389\\
4,3;1&1.978&-1.740&3.708 & -2.024& 4.663& -3.708\\
2,1;2&1.945&-1.572&3.676 & -2.054& \multicolumn{1}{c}{$\cdots$} &
\multicolumn{1}{c}{$\cdots$} \\
2,2;2&1.967&-1.682&3.641& -1.987& 7.715& -3.531\\
2,3;2&2.008&-1.900&3.799& -2.269& 7.277& -2.916\\
3,2;2&\multicolumn{1}{c}{$\cdots$}&\multicolumn{1}{c}{$\cdots$}&3.874&
-2.468&7.250& -2.899\\
3,3;2&1.971&-1.679&3.773& -2.210& 7.308& -2.920\\
2,1;3&1.959&-1.628&3.599& -1.889& 6.802& -2.680\\
2,2;3&1.984&-1.784&3.747& -1.956& 7.508& -3.404\\
2,3;3&1.982&-1.778&3.777& -2.110& 7.003& -2.539\\
3,2;3&1.981&-1.770&3.779&
-2.034&\multicolumn{1}{c}{$\cdots$}&\multicolumn{1}{c}{$\cdots$} \\[0.8ex]
\tableline
\end{tabular}
\normalsize
\end{table}

\begin{table}[!hbt]
\centering
\caption[Approximate position of singularities with exponents]{\centering Approximate position of singularities with exponents}
\label{tab:diff2}
\vspace{2ex}
\begin{tabular}{llll}
\tableline
\tableline
\\[-1.5ex]
\multicolumn{1}{c}{$D$}& \multicolumn{1}{c}{$B_2\rho_{sing}$} &\multicolumn{1}{c}{$\eta_{sing}$} &
\multicolumn{1}{c}{$\phi$}\\[0.7ex]
\tableline
\\[-1.5ex]
1 & 1.00 & 1.00 & -1.00\\
2 & 1.98 & 0.99 & -1.75\\
3 & 3.75 & 0.94 & -2.1 \\
4 & 7.00 &0.88 & -3\\[0.8ex]
\tableline
\end{tabular}
\normalsize
\end{table}

\section{Conclusion}
\label{conclusion}

In \tab{tab:numericalvirial} above   we have reported
the first computations of
the virial coefficients $B_9$ and $B_{10}$ for hard spheres in
dimensions $D=2,\cdots, 8$, and shown that $B_8$ is negative for $D \ge
5$. The coefficient $B_{10}$
is negative for $D>4$ and at $D=4$ the ratios of sucessive
coefficients  oscillate in such a way as to suggest that negative
values may occur for $k=12-14.$ For $D=2,3$ analysis of the first 10
virial coefficients leads to a radius of convergence greater than
close packing. This is in agreement with conclusions reached in
previous studies based on 8 or fewer virial coefficients. The meaning
of this is controversial and the present authors argue that the true
large $k$ behaviour is not seen in the first 10 coefficients. More
complete analysis and discussion of this is given in \cite{clisby2004d}.

\vspace{0.2cm}

\noindent {\bf Acknowledgments:} {This work was supported in part by
the National Science Foundation under DMR-0302758.}

\end{document}